# Extreme Loading of Aircraft Fan Blade

Dibakar Datta

Erasmus MSc in Computational Mechanics

International Center for the Numerical Method in Engineering (CIMNE)

Universitat Politecnica de Catalunya(UPC)

Campus Nord, Barcelona, Spain

Present Address: dibakar_datta@brown.edu or dibdatlab@gmail.com

**Abstract:** *The response of an aircraft fan blade manufactured by composites under the action of static and impact load has been studied in this report. The modeling and analysis of the geometry has been done using CASTEM 2007 version. For the quasi static analysis, the pressure has been incrementally applied until it satisfies the failure criteria. The deformed configuration, strain, Von-Mises stress, and the deflection of the blade have been studied. The response of the system e.g. deformation time history due to the impact of the projectile has been studied where the Newmark method for the dynamic problem has been implemented.*

1.  **Introduction.**

With the advancement in the development of composite materials in recent past, there are now plenty of possibilities to implement these structures into highly strained components application. Hence in aircraft industries, composites are now widely used (e.g. composite fan blade). The main aim is to create lightweight structures with an extraordinary high utilization of mechanical strength and fatigue life. Future transport aircraft will contain composites for primary and secondary wing and fuselage components.

However, composites such as carbon fibre/epoxy are inherently brittle and usually exhibit a linear elastic response up to failure with little or no plasticity. Hence composite structures are vulnerable to impact damage and have to satisfy certification procedures for high velocity impact from runway debris or bird strike. When suitably triggered to fail by delimitation and compression crushing, composites may exhibit high energy absorption and are of interest for light weight energy absorbing structural elements such as subfloors in helicopters and transport aircraft. Thus for the further development of composite aircraft structures, especially in safety critical components, it is important to understand the mechanisms of energy absorption and failure, and to have predictive design tools for simulating the response of composite structures under impact and crash loads.

Lots of studies have been done on this topic. *Haug et.el. (1993), Ladeveze et.el(1992), Jhonson et.el.(1998)* studied the crash response of the composite structures. *Curiel et.el.(2006)* studied about improved predictive modeling of strain localization and ductile fracture of composite subjected to impact. One of the major issues regarding composites is the failure mechanism of composites. *Christensen et. el.(1997), Hashin et.el(1980), Tsai et.el.(1971)*, have addressed the problem of predicting correct failure mechanism of composites.

In this project, an aircraft fan blade manufactured by carbon reinforced epoxy has been considered for the analysis of the response of the composites under static and dynamic loading.

## 2. Problem Description:

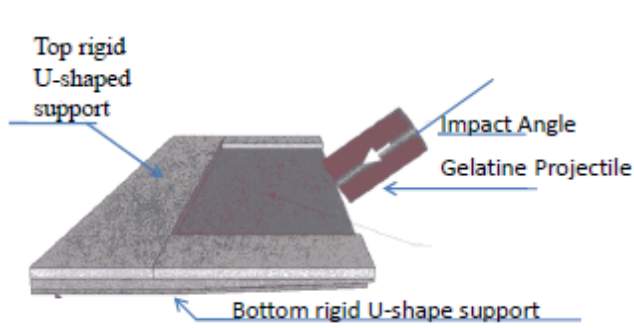
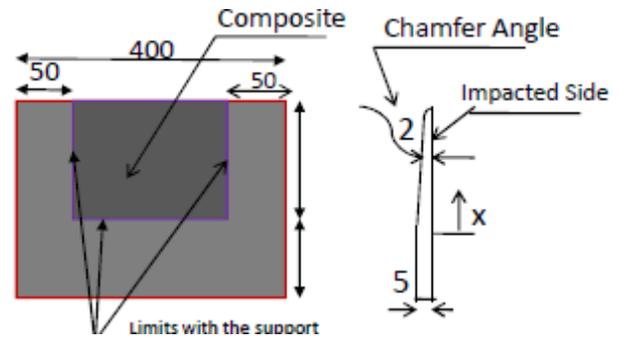

Fig . Configuration of the model          Fig . Dimension of the model

**Analysis:**

1. *Quasi-static loading:* **A**. Increment of pressure until collapse and Plotting of deformed configuration.

    **B**. Maximum effective strain reached in the targeted area.

    **C**. Von-Mises stress field.   **D**. Maximum deflection at the center of the specimen.

2. *Impact Loading:* **A.** Implementation of dynamic analysis using Newmark.  **B**. Plot of impact load /pressure v/s time evolution. **C.** Deformed configuration under the impact with velocity 475 m/s

## 3. Analysis for Quasi-static loading:

### 3.1 Increment of pressure until collapse and Plotting of deformed configuration:

The pressure at the time of collapse of the structure is not known 'a priori'. Hence the static analysis is performed starting with a lower value of pressure and incrementing it until the failure criteria is satisfied. The pseudo code is given below:

***Pseudo Code.1***: *Implementation of quasi-static loading*

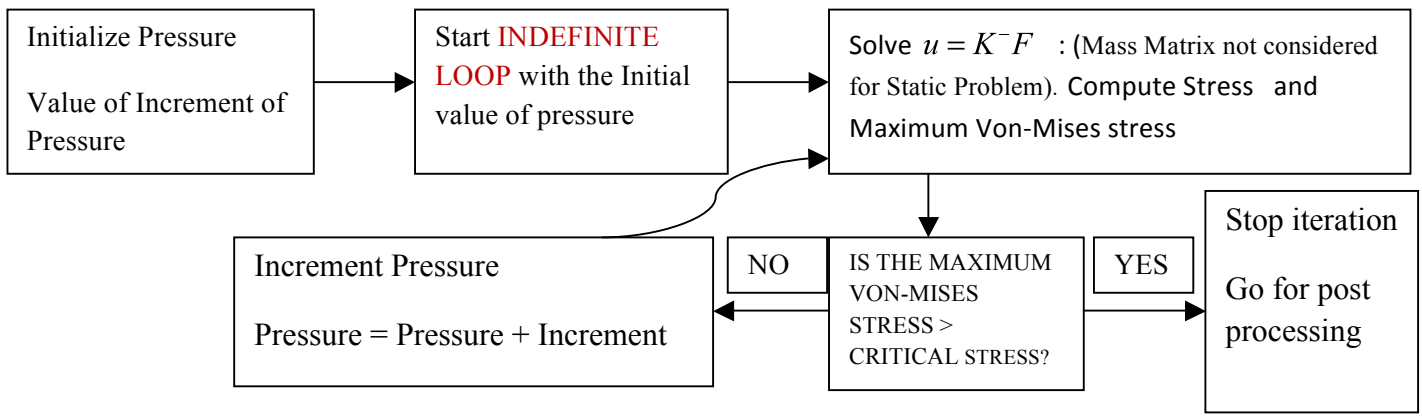

***Failure Criteria***: The value of the collapse load depends on the choice of the failure criteria. In recent past, several failure criteria have been proposed like Matrix Controlled Failure Criteria, Fiber Controlled Failure, Tensile Matrix Mode Failure, Compression Matrix Mode Failure, Tensile Fiber Mode Failure, Compression Fiber Mode Failure, Tsai-Wu criteria for failure etc. However, for this project the failure criteria has been chosen as: $\sigma_{VM} - \sigma_y > 0$ ;

$\sigma_{VM}$ : Maximum Von-Mises stress

$\sigma_y$ : Average of the fibre normal strength along its direction and perpendicular direction. $= \dfrac{1950 + 48}{2} = 999 \simeq 1000$ MPa

***Deformed Configuration:***

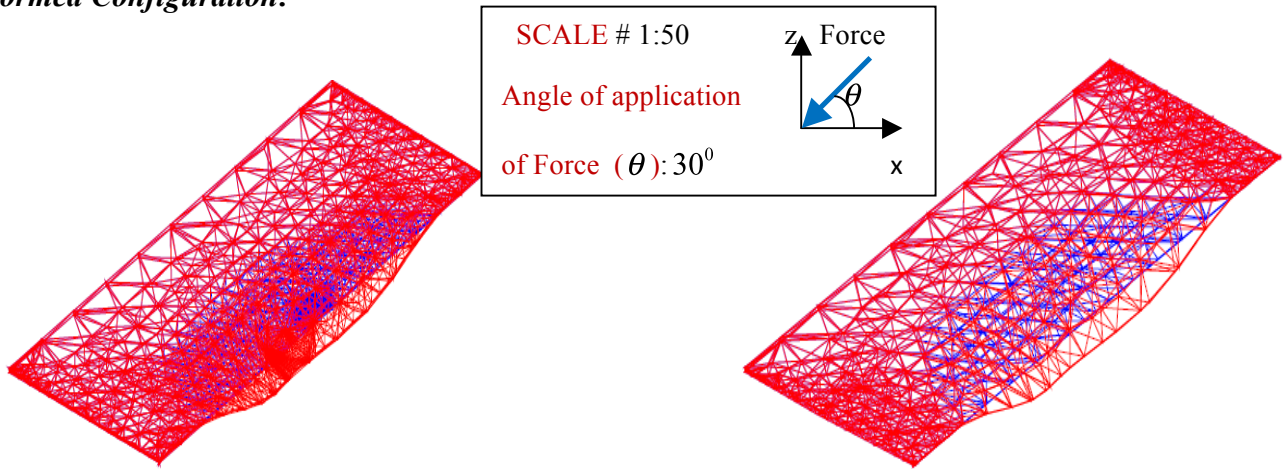

SCALE # 1:50

Angle of application of Force ($\theta$): $30^0$

Fig : Deformed shape of the *Upper Layer*     Fig : Deformed shape of the *Lower Layer*

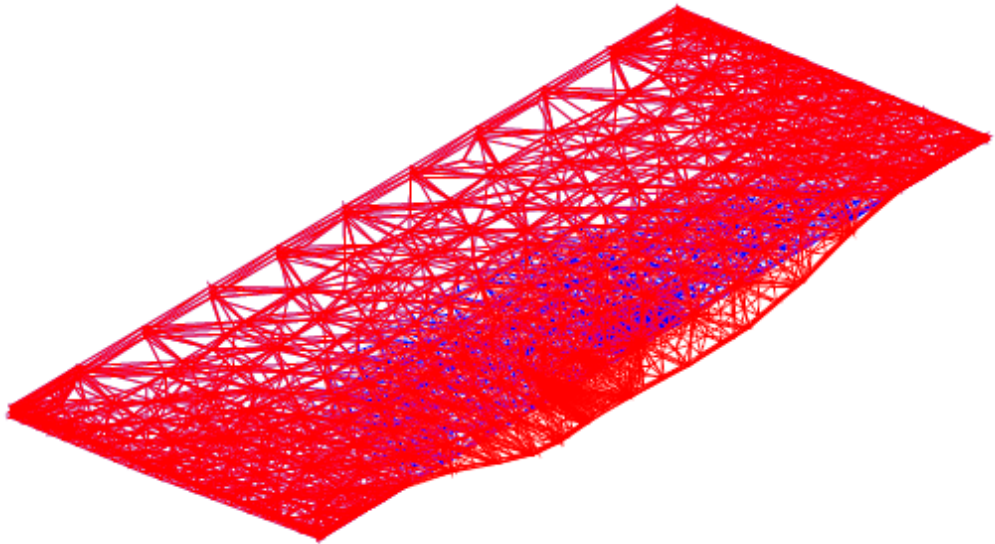

Fig : Deformed shape of the *Whole System (The fan blade)*

Legend:

— Undeformed Shape

— Deformed Shape

Observation:

- The magnitude of the collapse load and hence the deformed shape differs considerably depending on the choice of failure criteria.

- The magnitude of collapse load also changes depending on the mesh and *even in different version of Cast3M!*

Collapse Load: The value of the collapse pressure P=200 MPa (Using Cast3M 2007)

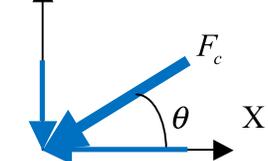

Fig : Applied Force with impact angle ($\theta$): $30^0$

Half circular zone of radius =19.50 mm where the load is applied.

$A = \pi \dfrac{(19.50)^2}{2} = 597.30$ mm

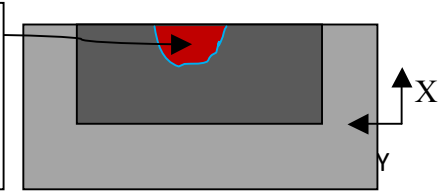

Fig : Half circular zone where the load is applied.

*Hence the collapse load= 200* 597.30 N = 119.46 kN*

## 3.2 Maximum effective plastic strain reached in the targeted area

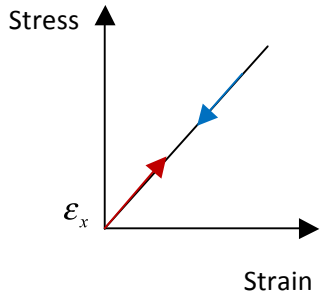
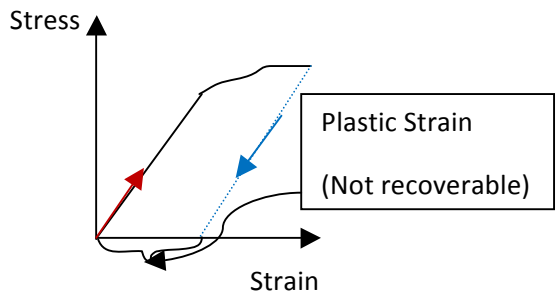

Fig : Stress-Strain curve for elastic material.

Fig :Stress-Strain curve for plastic material.

Plastic Strain (Not recoverable)

LEGEND

→ Loading    ← Unloading

NOTE:

For this problem the material used is

MECANIQUE ELASTIQUE ORTHOTROPE

Hence there will not be any plastic Strain. Strains at collapse are computed.

*Table: Maximum Strain in the targeted area:*

| | Normal Strain | | | Shear Strain | |
|---|---|---|---|---|---|
| | Maximum -ve | Maximum +ve | | Maximum -ve | Maximum +ve |
| $\varepsilon_x$ | -1.75E-02 | 1.35E-02 | $\gamma_{xy}$ | -1.58E-02 | 9.45E-03 |
| $\varepsilon_y$ | -3.16E-03 | 3.81E-03 | $\gamma_{yz}$ | -3.7E-03 | 3.38E-03 |
| $\varepsilon_z$ | -2.92E-03 | 5.41E-03 | $\gamma_{zx}$ | -1.4E-02 | 7.66E-03 |

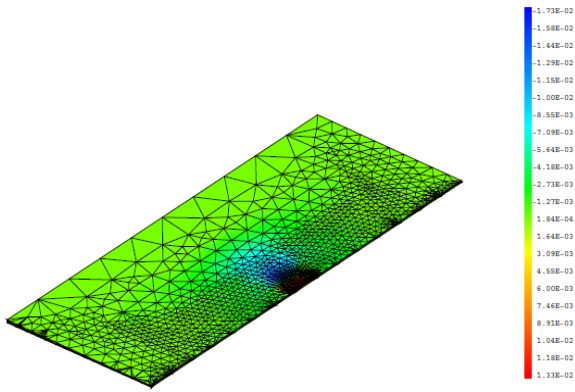
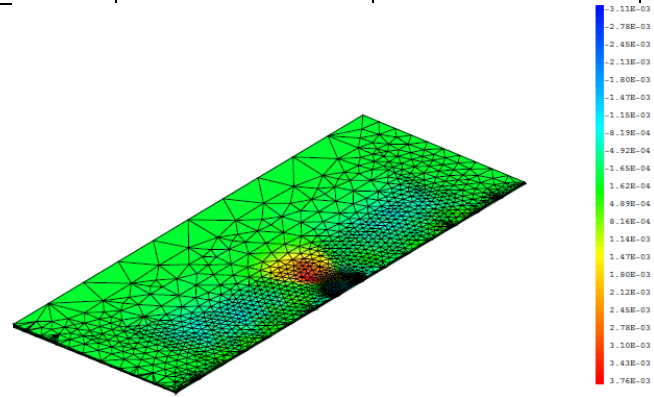

Fig : $\varepsilon_x$ field in the upper layer

Fig : $\varepsilon_y$ field in the upper layer

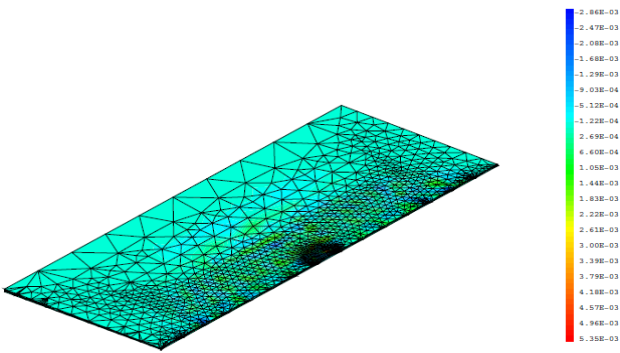
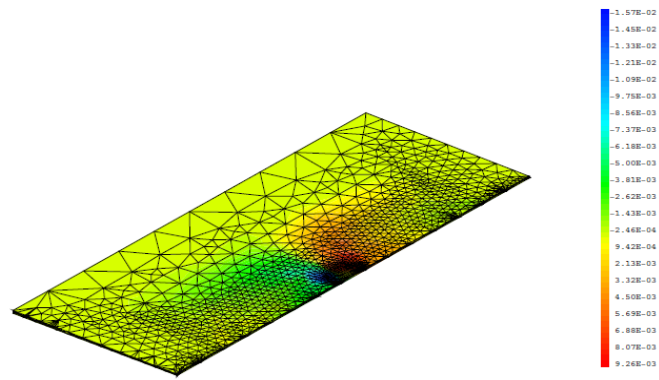

Fig : $\varepsilon_z$ field in the upper layer

Fig : $\gamma_{xy}$ field in the upper layer

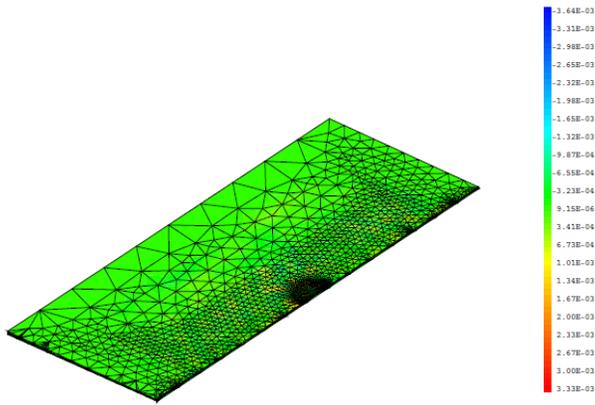
Fig : $\gamma_{yz}$ field in the upper layer

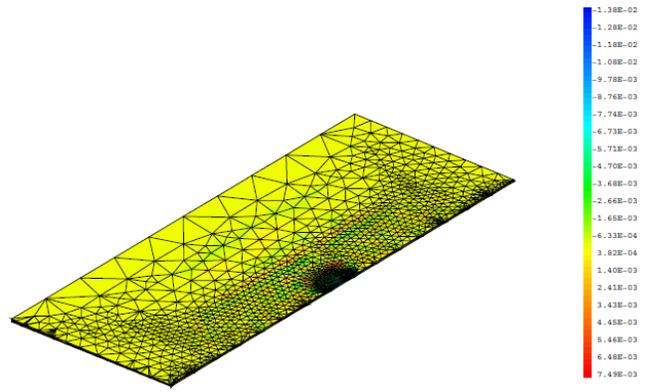
Fig : $\gamma_{zx}$ field in the upper layer

### 3.3 Von-Mises Stress Field:

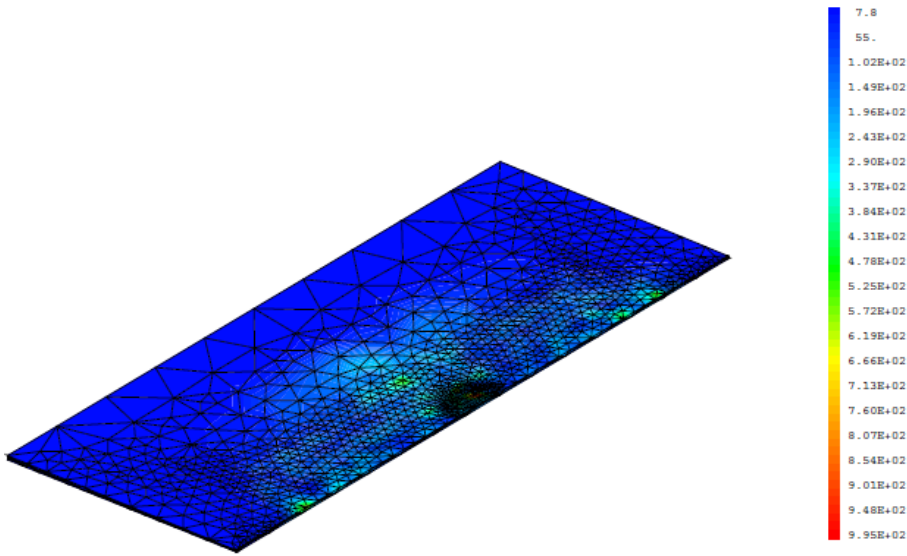
Fig : Von-Mises Stress Field in Upper-Layer

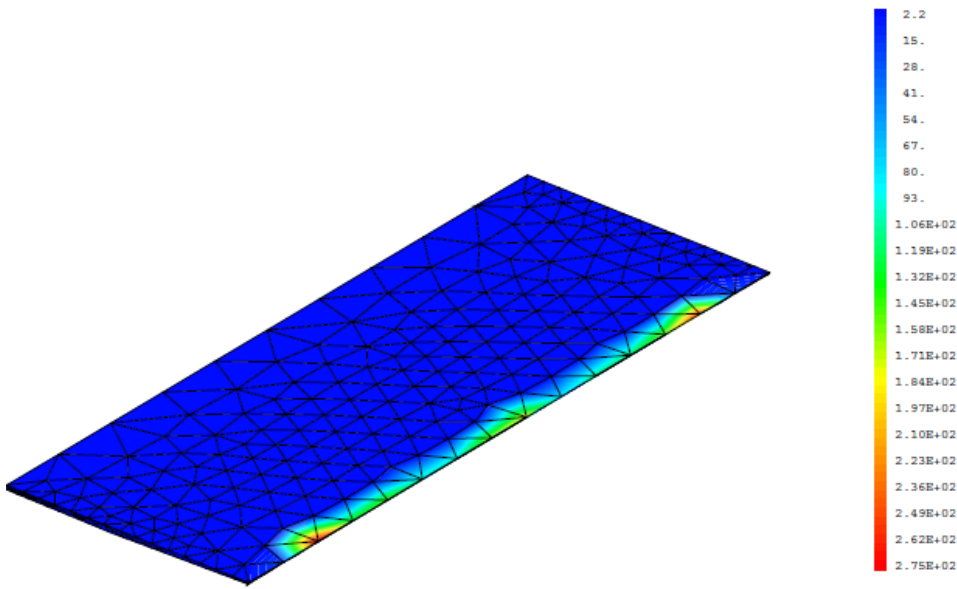
Fig : Von-Mises Stress Field in Lower-Layer

NOTE:

- The Von-Mises stress in the upper layer is much higher than that in the lower layer. Because the lower layer is not subjected to loading. When Von-Mises stress in the upper layer is exceeding 1000MPa that in the lower layer is only 166MPa.

- Von-Mises stress field gives an idea where the crack will initiate in real life when the model is subjected to loading. As it can be observed that crack will initiate at the zone of loading as well as at the edges of the support.

- As it can be observed, there will not be any considerable damage at the time of collapse of the fan blade in the lower layer.

## 3.4 Maximum Deflection in the center of the specimen:

| Displacement (mm) | Ux | Uy | Uz |
|---|---|---|---|
| Maximum (Using Cast3M 2007) | -0.46 | -1.3E-03 | -0.805 |

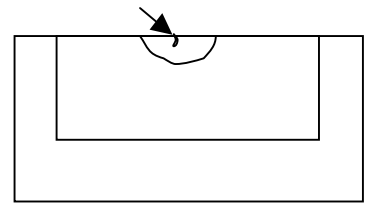

Point considered

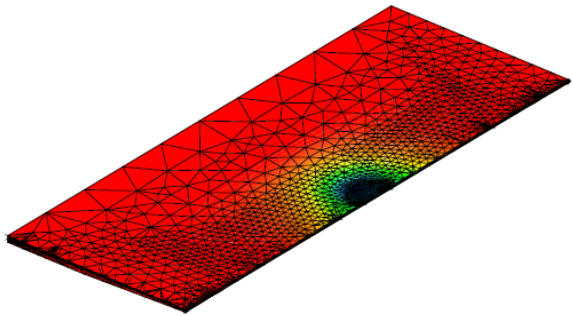

Fig: Displacement field along X direction

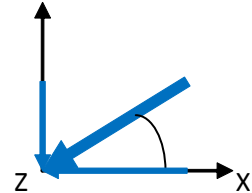

NOTE:

- Horizontal force is causing *compression* on the fan blade. Hence x-displacement at each fiber point is negative.

- The vertical force is causing *bending* of the fiber. Thu z displacements are all negative.

- Displacement along Y direction is because of the *Poisson effect* of the material. Since the loading is applied on the axis of symmetry, the Y direction displacement is symmetric w.r.t the plan of loading.

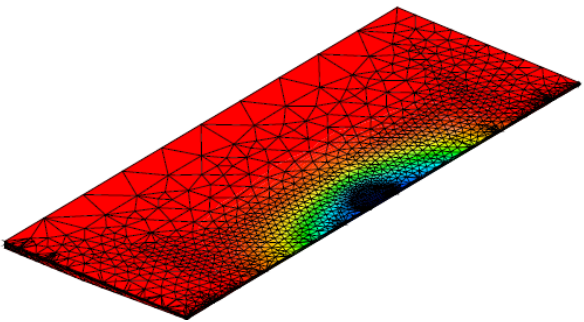

Fig: Displacement field along Z direction

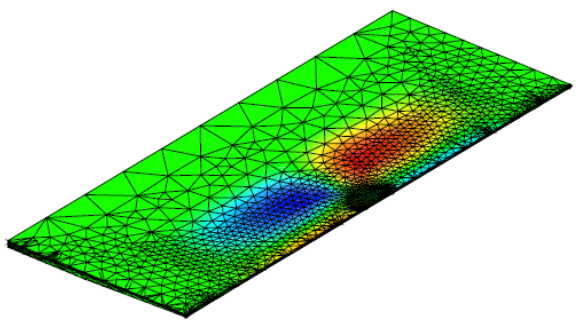

Fig: Displacement field along Y direction

# 4. Analysis for Impact Loading

The governing equation of the system is given by:

$M\ddot{X} + C\dot{X} + KX = f(t)$, Where $f(t) = F_{impact}\delta(t)$, $\delta(t)$ is the Dirac Delta

The damping has not been modeled in this problem, Hence C=[0].

## Modeling Dirac Delta:

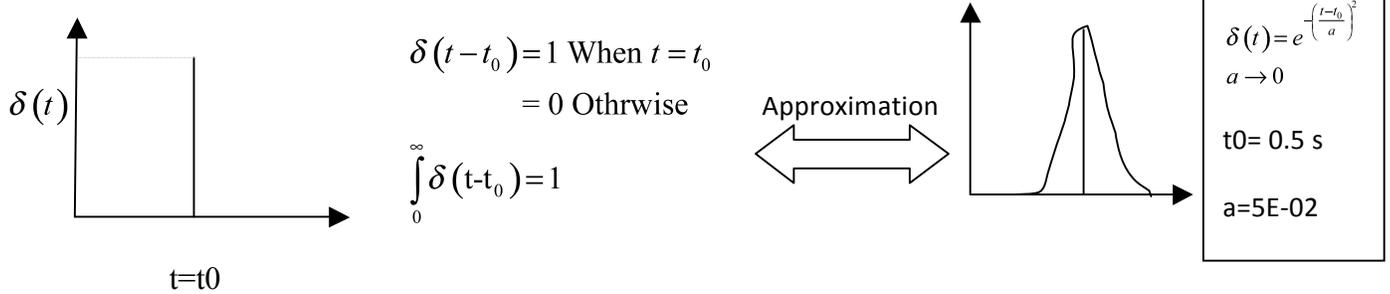

$\delta(t-t_0) = 1$ When $t = t_0$
$= 0$ Othrwise
$\int_0^\infty \delta(t-t_0) = 1$

Approximation

$\delta(t) = e^{-\left(\frac{t-t_0}{a}\right)^2}$
$a \to 0$
t0= 0.5 s
a=5E-02

Fig: Modeling Dirac Delta function with exponenetial function.

## Calculation of Impact Load:

Impact load = (Change in Momentum/ Time of impact). Since after the impact, the velocity of the projectile is zero, hence change in momentum = (MV-0). Where, M= Mass of the Projectile, V=Velocity of the Projectile

Change in Momentum: $MV$. $M = \rho Vol = (0.96 \times 10^{-9} \times \pi \times (19.5)^2 \times 68.5) = 7.856 \times 10^{-5}$ kg; $Vel = 475 \times 10^3$ mm/s

Time of Impact: $\Delta t = \frac{L}{v_s}$; L = Length of the projectile ; $v_s$ = Speed of the sound through the projectile $\left(= \sqrt{\frac{E}{\rho}}\right)$

$E = 2G(1+v) = 10.40$ MPa, G(Shear Modulus) $= 4 MPa$, $v$(Poission Ratio) $= 0.30$(Assumed)

$v_s = \sqrt{\frac{10.40}{0.96 \times 10^{-9}}} \left(\frac{N.mm^3}{mm^2.kg}\right)^{\left(\frac{1}{2}\right)} = 3.3 \times 10^6 \frac{mm}{s}$; $\Delta t = \frac{L}{v_s} = \frac{68.50}{3.3 \times 10^6} s = 2.08 \times 10^{-5} s$

$F = \frac{MV}{\Delta t} = \frac{7.856 \times 10^{-5} \times 475 \times 10^3}{2.08 \times 10^{-5}} \frac{kg.mm}{s^2} = 1793.95 \frac{kg.m}{s^2} \simeq 1800$ N

Hence Applied Pressure($P_{impact}$) $= \frac{F}{A} = \frac{1800 \times 0.50}{\pi \times (19.50)^2} = 3.013$ Mpa

## 5.1 Configuration and Implementation of the dynamic analysis using Newmark.

As mentioned before, the linear dynamic equilibrium equation can be written as:

$M\ddot{u}_t + C\dot{u}_t + Ku_t = F_t$

Using Taylor series, we get.

$u_t = u_{t-\Delta t} + \Delta t \dot{u}_{t-\Delta t} + \frac{\Delta t^2}{2} \ddot{u}_{t-\Delta t} + \frac{\Delta t^3}{6} \dddot{u}_{t-\Delta t} + \ldots$

$\dot{u}_t = \dot{u}_{t-\Delta t} + \Delta t \ddot{u}_{t-\Delta t} + \frac{\Delta t^2}{2} \dddot{u}_{t-\Delta t} + \ldots$

Truncating these equations, we get.

$$u_t = u_{t-\Delta t} + \Delta t \dot{u}_{t-\Delta t} + \frac{\Delta t^2}{2}\ddot{u}_{t-\Delta t} + \beta \Delta t^3 \dddot{u}_{t-\Delta t}$$

$$\dot{u}_t = \dot{u}_{t-\Delta t} + \Delta t \ddot{u}_{t-\Delta t} + \gamma \Delta t^2 \dddot{u}_{t-\Delta t}$$

Assuming acceleration to be linear within the time step, we get $\dddot{u} = \frac{(\ddot{u}_t - \ddot{u}_{t-\Delta t})}{\Delta t}$. Substituting this equation in the previous two equations, we get:

$$u_t = u_{t-\Delta t} + \Delta t \dot{u}_{t-\Delta t} + \left(\frac{1}{2} - \beta\right)\Delta t^2 \ddot{u}_{t-\Delta t} + \beta \Delta t^3 \dddot{u}_{t-\Delta t}$$

$$\dot{u}_t = \dot{u}_{t-\Delta t} + (1-\gamma)\Delta t \ddot{u}_{t-\Delta t} + \gamma \Delta t^2 \dddot{u}_{t-\Delta t}$$

These two equations together with the governing equation of motion can be iteratively solved for each time step for each displacement DOF of the structural system. The term $\ddot{u}_t$ is obtained from the governing equation by dividing the equation by the mass associated with the DOF.

The command DYNAMIC in Cast3M implements this algorithm. User has to provide a table for initial velocity

Displacement, and rigidity, mass matrix, loading, frequency, time interval.

**5.2 Plotting of the impact pressure time evolution.**

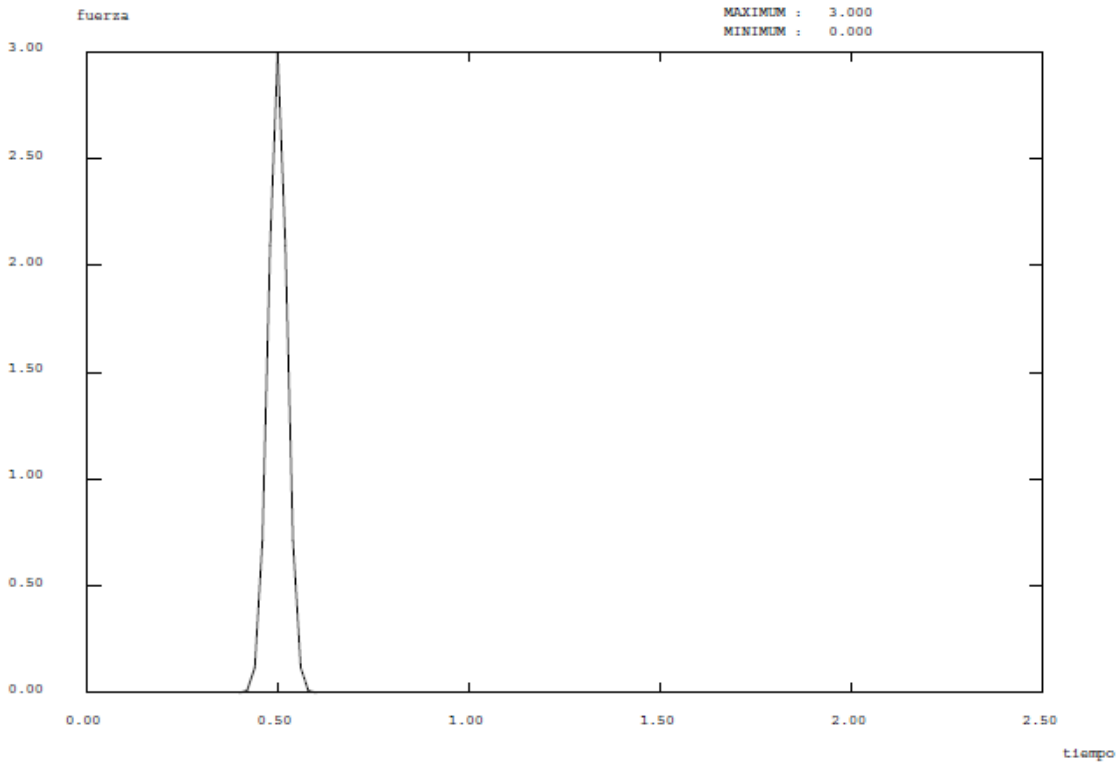

Fig. Impact pressure time evolution.

(As shown in the calculation, impact pressure=3Mpa. The impule is modeled using Dirac Delta approximated by exponential function. Hence equivalent pressure due to impact is 3Mpa is suddenly applied at time t=0.50 s for an infinitesimal period of time. )

### 5.3 Deformed configuration when the aerofoil is impacted with the given projectile at 475 m/s

(The impulse is assumed to have been applied at t=0.5s) (SCALE : 1:50000)

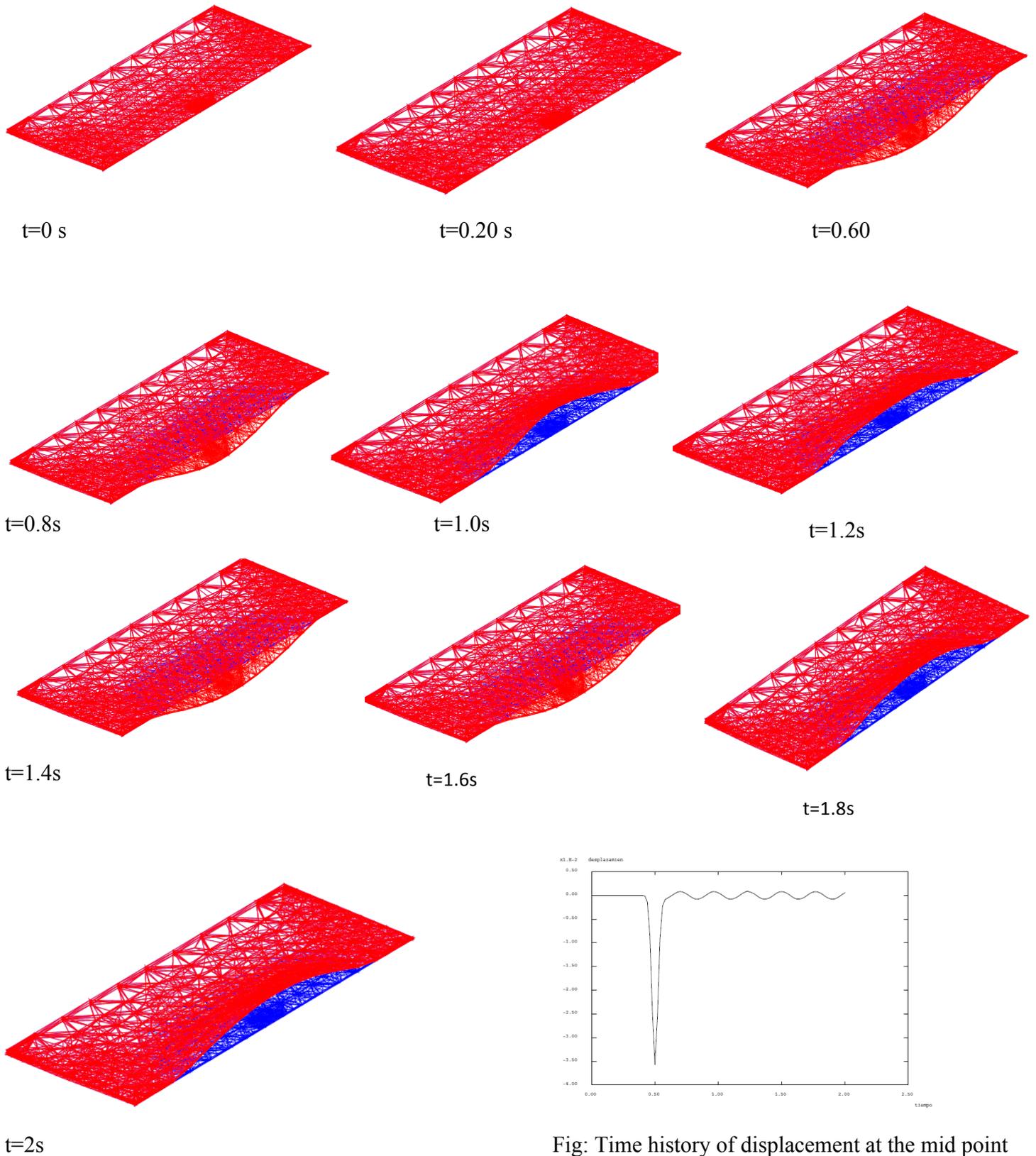

t=0 s

t=0.20 s

t=0.60

t=0.8s

t=1.0s

t=1.2s

t=1.4s

t=1.6s

t=1.8s

t=2s

Fig: Time history of displacement at the mid point along the line of the loading zone.

NOTE: There is **no damping** in the system. Hence the oscillation is not decaying.

# 6. Conclusions

- **Use of** Cast3M: The numerical analysis has been carried out using cast3M 2007.With the change in the mesh size and even using different version of Cast3m, the solution differs considerably.

- **Mesh Refinement:** Mesh should be very dense at the location where the loading is applied. Mesh size at the far boundaries is not so important.

- **Boundary condition:** If instead of modeling the U-shape support, only the flexible part is modeled and its edges are fixed, it will NOT properly addressed the physics of the problem.

- **Failure Criteria:** Choice of failure criteria is very important as different failure criteria will yield different result.

- **Mesh Type:** Solution may change depending on the choice of the mesh type. Mesh type should be such that it can capture the sharp edges and does not get distorted.

- **Orientation of the layer:** In this problem, two layers of fibers has been used in two orthotropic direction. Orientation of layers in different direction may yield different result.

- **Location of load application:** The area of the load applied is assumed in this case to be the half circular area. In reality, this may not be the case.

- **Impact loading:** Impact loading has been calculated assuming 100% elastic impact. In practice this may not be the case. Hence, the dissipation of energy and in elastic impact needs to be considered.

- **Importance of Impact analysis:** Usually the composite materials exhibit elastic response untills failure with little or no plasticity. Hence impact analysis is very important. In this case, damping is not modeled. However, to understand real dynamic behavior of composite, damping needs to be modeled.

- **Newmark Method:** Cast3M uses Newmark method for the dynamic analysis. However, for certain frequency range the method is unstable. Hence if the natural frequency of the system falls in that range, newmark method cannot be applied.

# References

1. Improved predictive modeling of strain localization and ductile fracture in a Ti-6Al-4V alloy subjected to

Impact loading, N Petrinic, J.L.Curiel Sosa, C.R.Siviour and B.C.F. Elliott, Journal de Physique IV (134):147-155(2006)

2. E. Haug, A. de Rouvray, *Crash response of composite structures*, Ch. 7 "Structural Crashworthiness and Failure" (ed) N. Jones and T. Wierzbicki, Elsevier, London (1993).

3. A.F. Johnson, D. Kohlgrüber, *Modelling the crash response of composite aircraft structures,* 8th European Conf. on Composite Materials (ECCM-8), Naples (1998).

## Acknowledgement:

The author of this project report is highly indebted to **Prof. J.L.Curiel Sosa,** the instructor of the computer modeling module for his outstanding guidance throughout the semester. It is only because of his soft guidance, the author was able to complete the project.